\documentclass[final,3p,times,twocolumn,authoryear]{elsarticle}
\usepackage{amssymb}
\usepackage{amsthm}

\journal{Journal of High Energy Astrophysics}

\begin{document}

\def\beq{\begin{equation}}
\def\enq{\end{equation}}
\def\ms{$M_{\odot}$}

\begin{frontmatter}



\title{Accretion Torque Reversals in GRO J1008-57 Revealed by {\em Insight-HXMT}}


\author[label1,label2]{W. Wang}
\author[label1,label2]{Y. M. Tang}
\author[label3,label4]{Y. L. Tuo}
\author[label1,label2]{P. R. Epili}
\author[label3,label4]{S. N. Zhang}
\author[label3]{L. M. Song}
\author[label3]{F. J. Lu}
\author[label3]{J. L. Qu}
\author[label3]{S. Zhang}
\author[label3]{M. Y. Ge}
\author[label3]{Y. Huang}
\author[label3]{B. Li}
\author[label3]{Q. C. Bu}
\author[label3]{C. Cai}
\author[label3]{X. L. Cao}
\author[label3]{Z. Chang}
\author[label5]{L. Chen}
\author[label3]{T. X. Chen}
\author[label6]{Y. B. Chen}
\author[label3]{Y. Chen}
\author[label3]{Y. P. Chen}
\author[label3]{W. W. Cui}
\author[label3]{Y. Y. Du}
\author[label3,label4]{G. H. Gao}
\author[label3,label4]{H. Gao}
\author[label3]{Y. D. Gu}
\author[label3]{J. Guan}
\author[label3,label4]{C. C. Guo}
\author[label3]{D. W. Han}
\author[label3]{J. Huo}
\author[label3]{S. M. Jia}
\author[label3]{W. C. Jiang}
\author[label3]{J. Jin}
\author[label3,label4]{L. D. Kong}
\author[label3]{C. K. Li}
\author[label3]{G. Li}
\author[label3,label4,label7]{T. P. Li}
\author[label3]{W. Li}
\author[label3]{X. Li}
\author[label3]{X. B. Li}
\author[label3]{X. F. Li}
\author[label3]{Z. W. Li}
\author[label3]{X. H. Liang}
\author[label3]{J. Y. Liao}
\author[label3]{B. S. Liu}
\author[label3]{C. Z. Liu}
\author[label3]{H. X. Liu}
\author[label3]{H. W. Liu}
\author[label3]{X. F. Lu}
\author[label3,label4]{Q. Luo}
\author[label3]{T. Luo}
\author[label3,label4]{R. C. Ma}
\author[label3]{X. Ma}
\author[label3]{B. Meng}
\author[label3,label4]{Y. Nang}
\author[label3]{J. Y. Nie}
\author[label3]{G. Ou}
\author[label3,label4]{X. Q. Ren}
\author[label3,label4]{N. Sai}
\author[label3]{X. Y. Song}
\author[label3]{L. Sun}
\author[label3]{L. Tao}
\author[label4]{C. Wang}
\author[label3]{L. J. Wang}
\author[label3,label4]{P. J. Wang}
\author[label8]{W. S. Wang}
\author[label3]{Y. S. Wang}
\author[label3]{X. Y. Wen}
\author[label3,label4]{B. Y. Wu}
\author[label3]{B. B. Wu}
\author[label3]{M. Wu}
\author[label3,label4]{G. C. Xiao}
\author[label3,label4]{S. Xiao}
\author[label3]{S. L. Xiong}
\author[label3,label4]{Y. P. Xu}
\author[label9]{R. J. Yang}
\author[label3]{S. Yang}
\author[label3]{J. J. Yang}
\author[label3]{Y. J. Yang}
\author[label3,label10]{B. B. Yi}
\author[label3]{Q. Q. Yin}
\author[label3]{Y. You}
\author[label3]{F. Zhang}
\author[label3]{H. M. Zhang}
\author[label3]{J. Zhang}
\author[label3]{P. Zhang}
\author[label3,label4]{W. Zhang}
\author[label3]{W. C. Zhang}
\author[label3]{Y. F. Zhang}
\author[label3,label4]{Y. H. Zhang}
\author[label3]{H. S. Zhao}
\author[label3,label4]{X. F. Zhao}
\author[label3]{S. J. Zheng}
\author[label3,label8]{Y. G. Zheng}
\author[label3,label4]{D. K. Zhou}

\address[label1]{School of Physics and Technology, Wuhan University, Wuhan 430072, China}
\address[label2]{WHU-NAOC Joint Center for Astronomy, Wuhan University, Wuhan 430072, China}
\address[label3]{Key Laboratory of Particle Astrophysics, Institute of High Energy Physics, Chinese Academy of Sciences, Beijing 100049, China}
\address[label4]{University of Chinese Academy of Sciences, Chinese Academy of Sciences, Beijing 100049, China}
\address[label5]{Department of Astronomy, Beijing Normal University, Beijing 100088, China}
\address[label6]{Department of Physics, Tsinghua University, Beijing 100084, China}
\address[label7]{Department of Astronomy, Tsinghua University, Beijing 100084, China}
\address[label8]{Computing Division, Institute of High Energy Physics, Chinese Academy of Sciences, Beijing 100049, China}
\address[label9]{College of physics Sciences and Technology, Hebei University, Baoding, Hebei Province 071002, China}
\address[label10]{School of Physics and Optoelectronics, Xiangtan University, Xiangtan, Hunan 411105, China}

\begin{abstract}
GRO J1008-57, as a Be/X-ray transient pulsar, is considered to have the highest magnetic field in known neutron star X-ray binary systems. Observational data of the X-ray outbursts in GRO J1008-57 from 2017 to 2020 were collected by the Insight-HXMT satellite. In this work, the spin period of the neutron star in GRO J1008-57 was determined to be about 93.28 seconds in August 2017, 93.22 seconds in February 2018, 93.25 seconds in June 2019 and 93.14 seconds in June 2020. GRO J1008-57 evolved in the spin-up process with a mean rate of $-(2.10\pm 0.05)\times$10$^{-4}$ s/d from 2009 -- 2018, and turned into a spin down process with a rate of $(6.7\pm 0.6)\times$10$^{-5}$ s/d from Feb 2018 to June 2019. During the type II outburst of 2020, GRO J1008-57 had the spin-up torque again. During the torque reversals, the pulse profiles and continuum X-ray spectra did not change significantly, and the cyclotron resonant scattering feature around 80 keV was only detected during the outbursts in 2017 and 2020. Based on the observed mean spin-up rate, we estimated the inner accretion disk radius in GRO J1008-57 (about 1 - 2 times of the Alfv\'{e}n radius) by comparing different accretion torque models of magnetic neutron stars. During the spin-down process, the magnetic torque should dominate over the matter accreting inflow torque, and we constrained the surface dipole magnetic field $B\geq 6\times 10^{12}$ G for the neutron star in GRO J1008-57, which is consistent with the magnetic field strength obtained by cyclotron line centroid energy.
\end{abstract}




\end{frontmatter}


\section{Introduction}
\label{Introduction}

GRO J1008-57 is a high-mass X-ray binary (HMXB) consisting of a neutron star and a Be star companion (Coe et al. 1994). GRO J1008-57 was discovered by the Burst and Transient Source Experiment (BASTE) on board the Compton Gamma-Ray Observatory (CGRO) in 1993. Pulsed flux in the 20--50 keV increased up to 1.1 Crab on July 18, 1993, and the spin period of $\sim$ 93.5 s was discovered (Stollberg et al. 1993). GRO J1008-57 is considered to have a highly eccentric orbit, $e$ = 0.68 $\pm$ 0.02, $a_{x}$sin$i$ = 530 $\pm$ 60 (light-second) , $P_{orb}$ = 247.8 $\pm$ 0.4 day (Coe et al. 2007), and the distance of the binary was estimated to be 5.8 $\pm$ 0.5 kpc (Riquelme et al. 2012). Detailed analysis on the 16-year RXTE/ASM light curve data of GRO J1008-57 discovered two modulation periods at $\sim 124.4$ d and 248.7 d (Wang 2014), suggesting that the neutron star passing the equatorial circumstellar disk of the Be star twice produces two X-ray flares in one orbital phase, near periastron (Type I outbursts) and apastron (mini flares).

GRO J1008-57 exhibits regular outbursts (Type I) due to accretion transfers during periastron passages as well as irregular giant (Type II) outbursts similar to other Be/X-ray pulsars. In last 15 years, Type II outbursts in GRO J1008-57 have occurred in June 2004, March, 2009, November 2012, November 2014, January 2015, August 2017, June 2020. Specially, outburst events in September 2014 (Type I, periastron outburst), November 2014 (type II outburst) and January 2015 (aphelion, Type II outburst) became the unique " Triple Peak" outbursts, and this "triple-peaked" outburst behavior had not been seen in any other source (K\"uhnel et al. 2017). The cause of the Type II outburst is currently unknown, needs to be studied furthermore (K\"uhnel et al. 2017). Moritani et al. (2013) suggested that in a highly centrifugal Be/X-ray binaries, if Be disk is not aligned with the orbital plane of the binaries, the neutron star can capture a large amount of matter during the event period, producing a Type II outbursts with high brightness when the misaligned Be disk passes through the neutron star's orbit.

%In the BATSE era, Bildsten et al. (1997) suggested the orbital period of $\sim 248$ d based on separation of three Type I outbursts. Shrader et al. (1999) found a period of $\sim 136.5$ d using %RXTE/ASM light curve data. Using the technique of pulse arrival time (Coe et al. 2007; K\"uhnel et al. 2013), they found the orbital period of $\sim 248$ d. Wang (2014) detailed analyzed the 16-year %RXTE/ASM light curve data of GRO J1008-57, and discovered two modulation periods at $\sim 124.4$ d and 248.7 d, suggesting that the neutron star passing the equatorial circumstellar disk of the Be star %near periastron (Type I outbursts) and apastron produces two X-ray flares.

Based on the BASTE data, Wilson et al. (1994) detected the pulse period between 20 keV and 160 keV, which was 93.548 $\pm$ 0.002s on 15 July 1993, 93.5665 $\pm$ 0.0005s on 23 July 1993 and 93.541 $\pm$ 0.004s on 10 August 1993. Shrader et al. (1999) reported a pulsed period of $\sim 93.62$ s in the 1993 outburst observed by ASCA. In 2007 November and December outbursts, pulsations with a period of 93.737 s were clearly detected in the light curves of the pulsar up to the 80 -- 100 keV energy band (Naik et al. 2011). During two outbursts in June 2004 and March 2009, the pulsation periods of $\sim$ 93.66s in 2004 and $\sim$ 93.73s in 2009 were determined by INTEGRAL observations (Wang 2014).  K\"uhnel et al. (2013)
derived the spin period at 93.679 s during the 2005 outburst, of 93.713 s during the 2007 outburst and 93.648 s during the 2012 outburst. Yamamoto et al. (2014) reported a period of 93.625 s in the 2012 outburst using Suzaku observations. Wang (2014) proposed that from 1993 until 2009, the spin period of GRO J1008-57 showed a downward trend with a rate $\sim$ 4 $\times$ 10$^{-5}$ s/d and after 2009, it may change to a spin-up trend. Recently, the Insight-HXMT collaborations reported a pulse period of $\sim 93.283$ s in the 2017 outburst (Ge et al. 2020).

It has been reported that the highest energy of cyclotron resonant scattering features (CRSFs) was measured in the hard X-ray spectrum of GRO J1008-57. In the early CGRO/OSSE measurements, a marginal feature around 88 keV (with only $2\sigma$) was reported (Grove et al. 1995). Yamamoto et al. (2014) reported that the 76 keV absorption characteristics (as the CRSF) were detected during the November 2012 outburst. Wang (2014) found that the CRSF energy of GRO J1008-57 was $\sim$ 74 keV in the 2009 outburst. Bellm et al. (2014) confirmed the 78 keV line feature ($\sim 4\sigma$) with the NuSTAR and Suzaku data. The Insight-HXMT collaborations (Ge et al. 2020) reported the CRSF with very high statistical significance ($>10 \sigma$) at a mean centroid energy of $\sim 90$ keV (with the line model {\em gabs} ) and $\sim 83$ keV (with the line model {\em cyclabs}) during the 2017 outburst.

China's first X-ray astronomical telescope satellite, the Hard X-ray Modulation Telescope (hereafter Insight-HXMT) was launched successfully in June 2017 (Zhang et al. 2020). Insight-HXMT has collected several X-ray observations on GRO J1008-57 from 2017 to 2020, covering four outbursts: two Type II outbursts in August 2017 and June 2020, Type I outbursts in February 2018 and June 2019. With more observations of Insight-HXMT, we can study the spin period evolution of GRO J1008-57 in last four years, and probe the spectral properties in different outbursts.

The observations and data analysis of Insight-HXMT on GRO J1008-57 were introduced in \S 2. In \S 3, the spin properties of the neutron star in GRO J1008-57 were studied, the torque reversals of the neutron star were revealed. In \S 4, the spectral properties in four outbursts will be shown for comparison. The conclusion and discussion are presented in \S 5.

\section{Observations and data analysis of Insight-HXMT}

Insight-HXMT has three main payloads: the High Energy X-ray Telescope (HE, Liu et al. 2020), the Medium Energy X-ray Telescope (ME, Cao et al. 2020), and the Low Energy X-ray Detector (LE, Chen et al. 2020).
\begin{itemize}
	\item[(HE)] NaI (CsI) detector has a range of 20--250 keV with the effective area of 5100 cm$^2$. The collimators of HE define 15 narrow field of view (FOV, 5.7$^{\circ}$ $\times$1.1 $^{\circ}$), 2 wide FOV(5.7$^{\circ}$ $\times$5.7$^{\circ}$ ) and a blind FOV which was covered with 2 mm tantalum.
	
	\item[(ME)] Si-PIN detector has a range of 5--30 keV with the effective area of 952 cm$^2$. ME consists of 3 detector boxes. Each box has 576 Si-Pin detector pixels read out by 18 ASIC (Application Specified Integrated Circuit). For each detector box, the collimators of ME confine 15 ASICs as narrow FOV(1$^{\circ}$ $\times$ 4 $^{\circ}$), 2 ASIC as wide FOV(4$^{\circ}$ $\times$ 4$^{\circ}$) and one blind FOV.
	
	\item[(LE)] SCD detectors range from 1 to 15 keV with the effective area of 384 cm$^2$. LE has three detection boxes. Twenty collimators have the small FOVs, 1.6$^{\circ}$ $\times$ 6$^{\circ}$ ; Six with the wide FOVs, 4$^{\circ}$ $\times$ 6$^{\circ}$ and two blind FOVs.
	
\end{itemize}

This work has used two data sets of P0114520, P0201012 (Observation ID).\footnote{Detailed introduction to data bases can be found at http://www.hxmt.org.} In Table 1, the description of the observed data by Insight-HXMT was presented.
% \usepackage{booktabs}
\begin{table}
	\caption{Insight-HXMT data used in this work.}
	\centering
\scriptsize
	\begin{tabular}{@{}ccccc@{}}
		\hline
		Obs. ID & Obs. date Start(UTC) & Duration (s) &  MJD &  \\ \hline
		P0114520001 & 2017-08-11 21:58:37.0 & 80789 & 57976 &  \\
		P0114520003 & 2017-08-18 11:30:22.0 & 212530 & 57983 &  \\
		P0114520005 & 2018-02-02 10:40:17.0 & 17589 & 58151 &  \\
		P0114520006 & 2018-02-09 00:12:19.0 & 17875 & 58158 &  \\
		P0114520007 & 2018-02-10 06:26:41.0 & 24004 & 58159 &  \\
		P0114520008 & 2018-02-12 22:06:24.0 & 17989 & 58161 &  \\
		P0114520009 & 2018-02-15 02:37:24.0 & 23432 & 58164 &  \\
		P0114520010 & 2018-02-16 19:59:42.0 & 17643 & 58165 &  \\
		P0114520011 & 2018-02-18 11:46:21.0 & 17641 & 58167 &  \\
		P0114520012 & 2018-02-21 11:21:55.0 & 17589 & 58170 &  \\
		P0114520014 & 2019-07-02 15:26:57.0 & 34667 & 58666 &  \\
		P0201012012 & 2019-06-22 05:25:13.0 & 35571 & 58656 &  \\
		P0201012014 & 2019-06-24 05:12:55.0 & 40372 & 58658 &  \\
		P0201012016 & 2019-06-26 16:13:46.0 & 23362 & 58660 &  \\
		P0201012018 & 2019-06-27 16:09:15.0 & 23332 & 58661 &  \\
		P0201012020 & 2019-06-29 14:18:31.0 & 36084 & 58663 &  \\
        P0201012355 & 2020-06-01 10:38:04.0 & 34534 & 59001 & \\
        P0201012356 & 2020-06-03 02:22:48.0 & 17431 & 59003 & \\
        P0201012357 & 2020-06-04 05:24:21.0 & 17433 & 59004 & \\
        P0201012358 & 2020-06-05 05:15:10.0 & 17644 & 59005 & \\
        P0201012359 & 2020-06-06 13:02:55.0 & 17624 & 59006 & \\
        P0201012360 & 2020-06-07 12:53:46.0 & 17622 & 59007 & \\
        P0201012361 & 2020-06-08 17:30:50.0 & 17622 & 59008 & \\
        P0201012362 & 2020-06-09 17:21:44.0 & 17623 & 59009 & \\
        P0201012363 & 2020-06-10 17:12:42.0 & 17621 & 59010 & \\
        P0201012364 & 2020-06-11 12:17:31.0 & 17621 & 59011 & \\
        P0201012365 & 2020-06-12 15:19:24.0 & 17621 & 59012 & \\
        P0201012366 & 2020-06-13 15:10:31.0 & 17622 & 59013 & \\
        P0201012367 & 2020-06-14 11:50:53.0 & 17621 & 59014 & \\
        P0201012368 & 2020-06-15 14:53:03.0 & 17619 & 59015 & \\
        P0201012369 & 2020-06-16 11:33:28.0 & 19622 & 59016 & \\
        P0201012440 & 2020-06-18 11:16:17.0 & 34803 & 59018 & \\
        P0201012441 & 2020-06-19 09:32:23.0 & 17624 & 59019 & \\
        P0201012442 & 2020-06-20 12:34:47.0 & 17626 & 59020 & \\
        P0201012443 & 2020-06-21 10:50:56.0 & 17625 & 59021 & \\
        P0201012444 & 2020-06-22 15:28:53.0 & 17625 & 59022 & \\
\hline
	\end{tabular}
\end{table}

The Insight-HXMT Data Analysis Software Package (HXMTDAS) V2.02 was used in this work. All science data of HE, ME and LE telescopes used small FOV detectors. We filtered the data with the following criteria: (1) pointing offset angle $< 0.1^\circ$; (2) pointing direction above Earth $> 10^\circ$; (3) geomagnetic cut-off rigidity value $> 8$; (4) time since SAA passage $> 300$ s and time to next SAA passage $> 300$ s; (5) for LE observations, pointing direction above bright Earth $> 30^\circ$. The methods of standard data reduction for the Insight-HXMT were introduced in previous publications (Huang et al. 2018; Xiao et al. 2020). Insight-HXMT has gone through a series of performance verification tests since the launch, at present shown the good calibration and estimation of the instrumental background (Li et al. 2020). Here we briefly summarized the data analysis procedures for the HXMTDAS V2.02.

\begin{itemize}
	\item[(i)] Calibration: Remove spike events caused by electronic system and calculate PI column values of event according to the Calibration Database (CALDB).
	\item[(ii)] Screening:
	\begin{itemize}
		\item[(a)] Generate a FITS file of good time interval (GTI).
		\item[(b)] Exclude some of the photons in event file then screen the data. The calibrated EVT file is filtered by applying cleaning criteria to produce a cleaned EVT file.
	\end{itemize}
	\item[(iii)] High level product extraction:
	\begin{itemize}
		\item[(a)] Extract spectra.
		\item[(b)] Extract light curves.
		\item[(c)] Generate the response files of energy spectra.
		\item[(d)] Generate background files of light curves/spectra.
		
	\end{itemize}
\end{itemize}

In the timing analysis, we have made the barycentric correction of the light curves using the tool {\em hxbary}. Based on the orbital ephemeris given by Coe et al. (2007), we made orbital motion correction. For the spectral analysis, we have used the energy bands as 3--10 keV (LE), 10--26 keV (ME) and 26-120 keV (HE) respectively, according to the present calibration results (Li et al. 2020).

\section{Spin period of the neutron star in GRO J1008-57}

\begin{figure}
	\includegraphics[width=9cm]{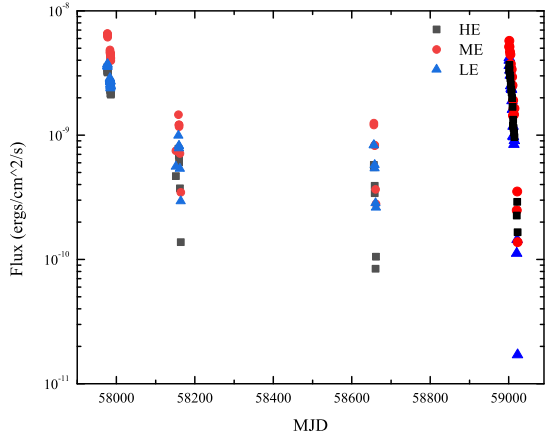}
	\caption{The flux variations of Be X-ray pulsar GRO J1008-57 from 2017 to 2020 determined by Insight-HXMT in three energy bands: 3--10 keV, 10 -- 26 keV, 26 -- 100 keV. The three observational epochs covered two type II outbursts in August 2017 and June 2020, and two type I outbursts in February 2018 and June 2019.
}
	
\end{figure}

The timing resolutions of Insight-HXMT can reach 2 $\mu$s (HE), 20 $\mu$s (ME), and 1 ms (LE), respectively. The temporal analysis here aimed to derive the rotation period of the neutron star in GRO J1008-57 for all observed data from 2017 -- 2020. In Fig. 1, we showed the flux variations of GRO J1008-57 from August 2017 to June 2020 in three energy bands: 3 -- 10 keV, 10 --26 keV, 26 -- 120 keV. The X-ray fluxes during two type I outbursts were much lower than those of two type II outbursts in 2017 and 2020.

We search for the periodical signal in a time series by folding data, determines the chi-square ($\chi^2$) of the folded light curve then plots the $\chi^2$ values versus the periods. The {\em efsearch} (a build-in function in HEAsoft) was used to complete the folding of the light curve and find the period. Using the {\em efsearch} task, we get the plots of $\chi^2$ values versus the periods. The true period can be defined as the position corresponding to the maximum value of the $\chi^2$. Therefore, we have used a Gaussian curve to fit the center position of the $\chi^2$ curves. In addition, {\em efsearch} might depend on the epoch (Li et al. 2012), so we corrected it as follows: calculate the $\chi^2$ with different epochs, $T(epoch) = t_0 + 0.1\,i\,P/N$, where $i$ = 0, 1, 2, ..., 9, and N is the number of phase bins in pulse profile. Then we averaged ten sets of $\chi^2$ versus period, fitted the peak with quadratic and find the maximum which should be the estimated period. The 1$\sigma$ error could be given when the quadratic is lowered by 1.0.

In the process of searching for the period, we have tried to obtain the more precise values with the high signal-to-noise observed data. In the type II outbursts in 2017 and 2020, we used the all data of LE, ME, HE detectors in the high flux levels, and the spin period values are basically same within the error of $\sim 0.001$ s. In this case, in order to get a higher quality period, HE, ME and LE files are used together here to reduce errors and find more accurate period. For the low flux levels, like the end of the 2007 outburst and type I outbursts in 2018 and 2019, we only used the detectors which have the better signal-to-noise light curve data, then obtained the spin period values.

In the Table 2, the spin period values of GRO J1008-57 in the four outbursts in August 2017, February 2018, June 2019 and June 2020 are presented. The variation of the spin period from 2017 -- 2020 is shown in Fig. 2. The neutron star in GRO J1008-57 showed a spin-up behavior from 2017 - 2018, while from 2018 - 2019, the accretion torque of the neutron star changed, and the star became to spin down. During the type II outburst in 2020, the neutron star transferred into the spin-up process again.
% \usepackage{booktabs}
\begin{table}
	\caption{Measurements of the spin period of the neutron star in GRO J2008-57.}
	\centering
\scriptsize
	\begin{tabular}{@{}lllll@{}}
		\hline
		MJD & Spin Period / s & Mission & Reference &  \\ \hline
		49182 & 93.5870 $\pm$ 0.0150 & BATSE & Stollberg et al. 1993 &  \\
		49205 & 93.6210 $\pm$ 0.0110 & ASCA & Shrader et al. 1999 &  \\
		50176 & 93.5500 $\pm$ 0.0200 & BATSE & Shrader et al. 1999 &  \\
		53166 & 93.6630 $\pm$ 0.0060 & INTEGRAL & Wang 2014 &  \\
		53168 & 93.6640 $\pm$ 0.0030 & INTEGRAL & Wang 2014 &  \\
		53171 & 93.6680 $\pm$ 0.0010 & INTEGRAL &Coe et al. 2007 &  \\
		53430 & 93.6793 $\pm$ 0.0002 & RXTE & K\"uhnel et al. 2013 &  \\
		54430 & 93.7134 $\pm$ 0.0002 & RXTE & K\"uhnel et al. 2013 &  \\
		54432 & 93.7370 $\pm$ 0.0010 & Suzaku & Naik et al. 2011 &  \\
		54912 & 93.7330 $\pm$ 0.0060 & INTEGRAL & Wang 2014 &  \\
		54920 & 93.7320 $\pm$ 0.0050 & INTEGRAL & Wang 2014 &  \\
		55658 & 93.7270 $\pm$ 0.0010 & RXTE & K\"uhnel et al. 2013 &  \\
		55917 & 93.7220 $\pm$ 0.0010 & Swift & K\"uhnel et al. 2013   &  \\
		56245 & 93.6480 $\pm$ 0.0020 & Swift & K\"uhnel et al. 2013   &  \\
		56251 & 93.6257 $\pm$ 0.0005 & Suzaku & Yamamoto et al. 2014 &  \\
		57976 & 93.2838 $\pm$ 0.0005 & HXMT/HE+ME+LE & this work &  \\
		57977 & 93.2868 $\pm$ 0.0010 & HXMT/HE+ME+LE & this work &  \\
		57983 & 93.2659 $\pm$ 0.0018 & HXMT/HE+ME+LE & this work &  \\
		57984 & 93.2666 $\pm$ 0.0004 & HXMT/HE+ME+LE & this work &  \\
		57985 & 93.2640 $\pm$ 0.0015 & HXMT/ME & this work &  \\
		58151 & 93.2117 $\pm$ 0.0024 & HXMT/HE & this work &  \\
		58158 & 93.2155 $\pm$ 0.0103 & HXMT/HE &this work  &  \\
		58159 & 93.2208 $\pm$ 0.0017 & HXMT/ME & this work &  \\
		58161 & 93.2206 $\pm$ 0.0012 & HXMT/ME &this work  &  \\
		58164 & 93.2186 $\pm$ 0.0106 & HXMT/ME & this work &  \\
		58165 & 93.2200 $\pm$ 0.0080 & HXMT/LE &this work  &  \\
		58167 & 93.2140 $\pm$ 0.0090 & HXMT/LE & this work &  \\
		58170 & 93.2260 $\pm$ 0.0010 & HXMT/HE & this work &  \\
		58656 & 93.2567 $\pm$ 0.0031 & HXMT/HE & this work &  \\
		58658 & 93.2625 $\pm$ 0.0085 & HXMT/HE & this work &  \\
		58660 & 93.2524 $\pm$ 0.0081 & HXMT/ME & this work &  \\
		58661 & 93.2568 $\pm$ 0.0053 & HXMT/ME &this work  &  \\
		58663 & 93.2560 $\pm$ 0.0160 & HXMT/LE & this work &  \\
		58666 & 93.2545 $\pm$ 0.0022 & HXMT/ME & this work &  \\
59002	&93.1445$\pm$ 	0.0011 & HXMT/HE+ME+LE & this work &  \\
59003	&93.1358$\pm$ 	0.0011 & HXMT/HE+ME+LE & this work &  \\
59004	&93.1352$\pm$ 	0.0013 & HXMT/HE+ME+LE & this work &  \\
59008&  93.1358$\pm$ 	0.0016  & HXMT/HE+ME+LE & this work &  \\
59010&	93.1285$\pm$ 	0.0021  & HXMT/HE+ME+LE & this work &  \\
59012&  93.1272$\pm$ 0.0018   & HXMT/HE+ME+LE & this work &  \\
59014&	93.1275$\pm$ 	0.0009  & HXMT/HE+ME+LE & this work &  \\
59016&	93.1198$\pm$ 	0.0039  & HXMT/HE+ME+LE & this work &  \\
59018&	93.1219$\pm$ 	0.0054  & HXMT/HE+ME+LE & this work &  \\
\hline
	\end{tabular}
\end{table}

In addition, we also have studied the evolution of pulse profiles of GRO J1008-57 in the different energy ranges from 2017 -- 2020. We used {\em efold} tasks to obtain pulse profiles in three energy bands: 3 --10 keV; 10 --26 keV; 26 -- 120 keV. The pulse profile of GRO J1008-57 had a double-peaked structure between 3--10 keV in all observations of four outbursts from 2017 -- 2020. Above 10 keV, the pulse profiles became a nearly single peak feature, but there are still the differences for the pulse features around the pulse phase 0.2 in four different outbursts (see Figures 3). In the type II outbursts of 2017 and 2020, there existed a mini second peak around the pulse phase 0.2 from 10-- 26 keV, and above 26 keV, the mini second peak was not obvious. For the two Type I outbursts in 2018 and 2019, the pulse profiles in the ranges of 10 --26 keV and 26 --120 keV showed the purely single peak features.




\begin{figure}
	\includegraphics[width=9cm]{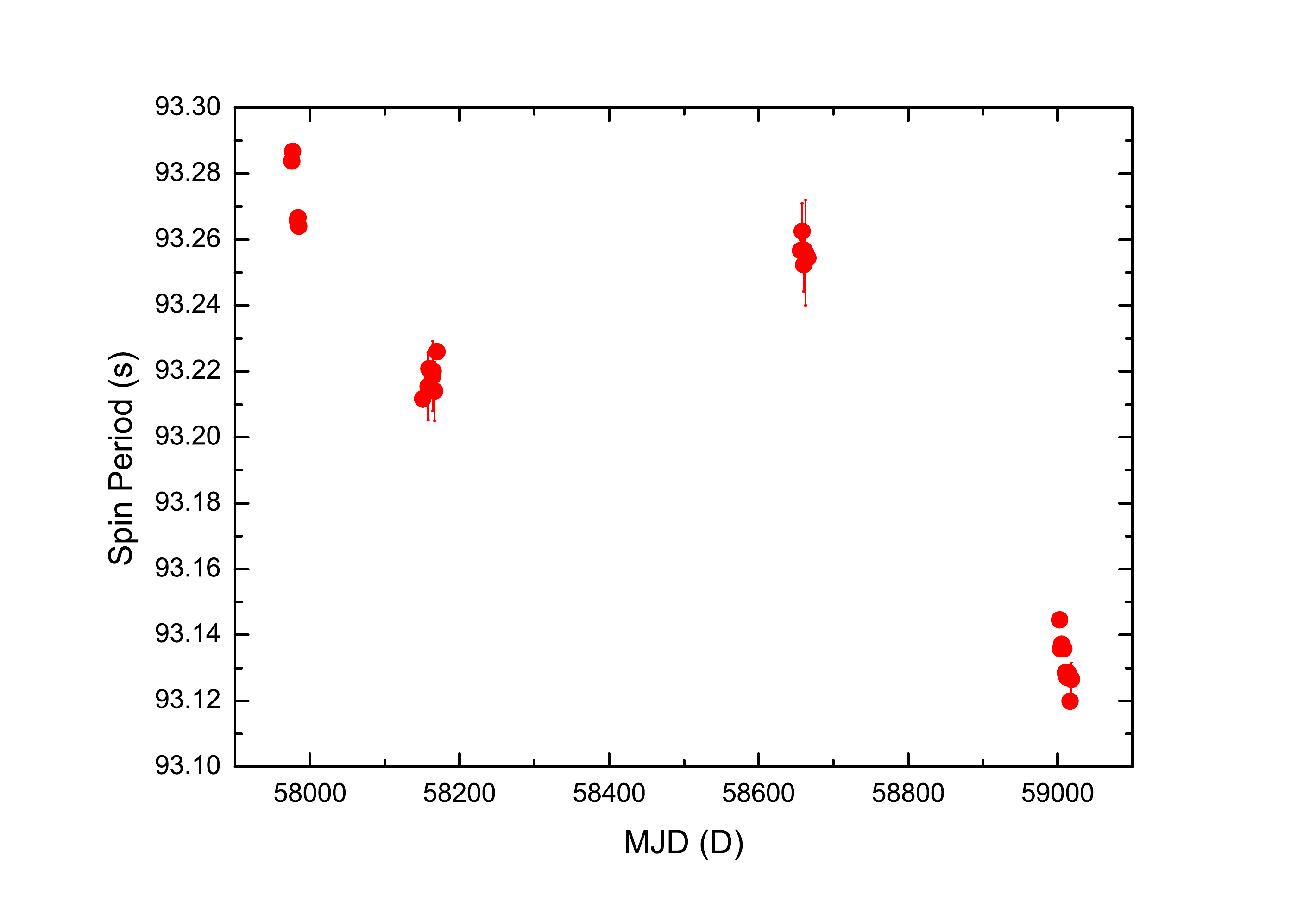}
	\caption{The spin period of the neutron star in GRO J1008-57 from 2017 to 2020 determined by Insight-HXMT. From August 2017 to Feb 2018, the neutron star continue to spin up, but after Feb 2018, the accretion torque of the neutron star changed and the neutron star became to spin down. During the outburst in 2020, GRO J1008-57 turned into the spin-up.}
	
\end{figure}

\begin{figure*}
\includegraphics[width=17cm]{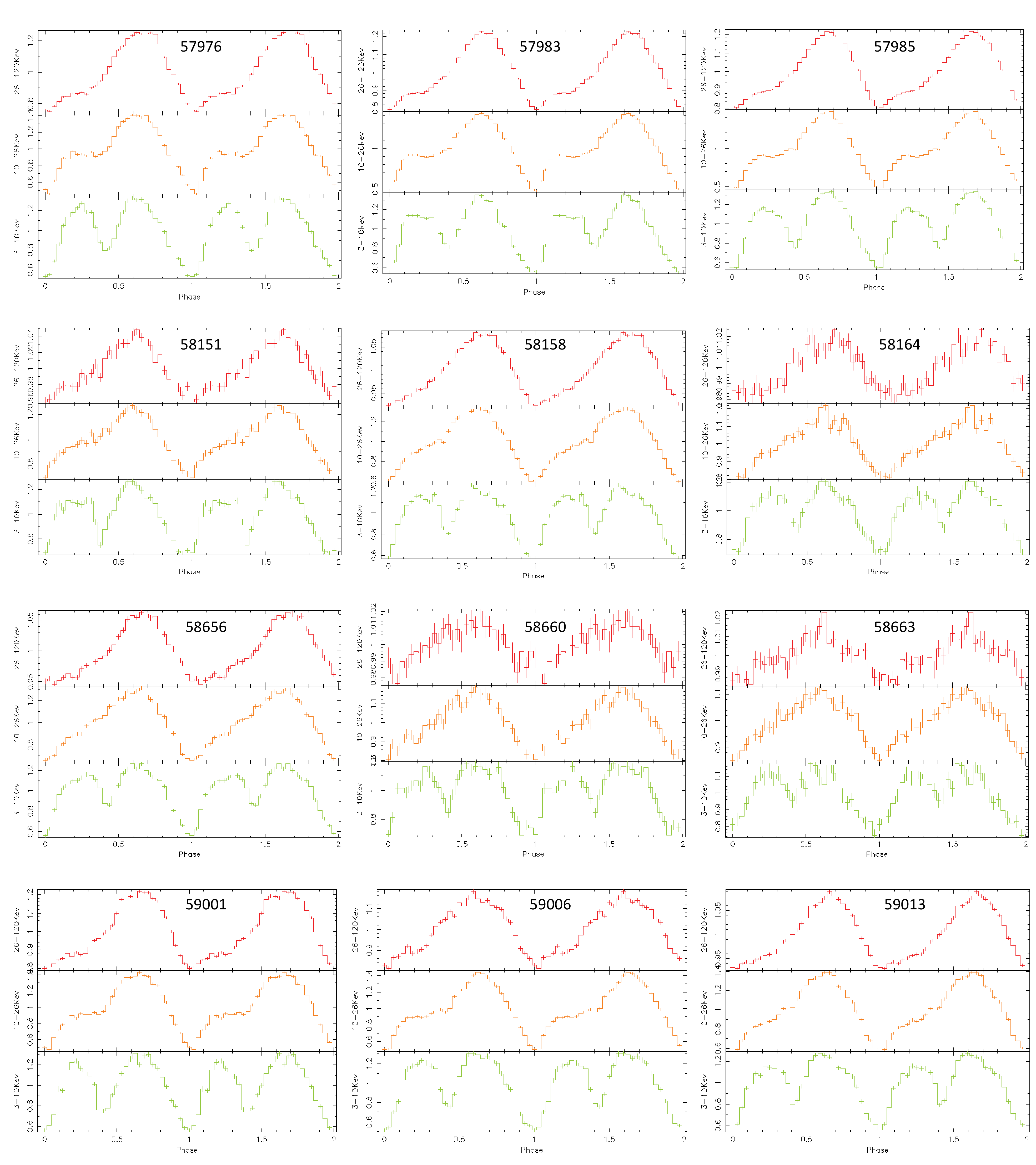}
\caption{Spin pulse profiles of GRO J1008-57 in different epochs of the four outbursts: (Top) MJD 57976, MJD 57983 and MJD 57985 in August 2017; (Middle-1) MJD 58151, MJD 58158, MJD 58164 in February 2018; (Middle-2) MJD 58656, MJD 58660, MJD 58663 in July 2019; (Bottom) MJD 59001, MJD 59006, MJD 59013 in June 2020. The pulse profiles are presented in three energy bands: 26--120 keV, 10--26 keV and 3--  10 keV.}
	
\end{figure*}

\section{Spectral properties of GRO J1008-57 in outbursts}

In this section, we will carry out the spectral analysis for GRO J1008-57 during the four outbursts, checking the possible spectral variations during the epochs of the accretion torque reversal. We have used the spectra from three detectors, covering the energy bands as 3 -- 10 keV (LE), 10 -- 26 keV (ME) and 26 -- 120 keV (HE) respectively. We have used the Xspec package version 12.10.1 in the following spectral analysis work. The X-ray spectrum of a neutron star X-ray binary can be generally described by a power-law model plus a high energy cutoff ({\em power*highecut}): $KE^{-\Gamma}$ when $E\leq E_{cut}$,  $KE^{-\Gamma}\exp-[ (E-E_{cut}) /E_{f}]$ when $E\geq E_{cut}$, where $\Gamma$ is the photon index of the power law, $E_{cut}$ is the cutoff energy in keV, $E_{f}$ is the e-folding energy in keV.  When we fit the spectra of GRO J1008-57 from 3--120 keV observed by Insight-HXMT, there existed the count excesses below 6 keV, we add the thermal component ({\em bbody}) to fit the soft X-ray band. We also tried other possible spectral models (i.e. {\em bmc, CompTT}) to fit the continuum spectra, however, the fittings were not good (reduced $\chi^2$ significantly larger 1). For the type I outbursts in 2018 and 2019, the spectra can be fitted with the continuum model, there were no significant absorption features in the range of 70 -- 90 keV. While for the type II outbursts in 2017 and 2020, after the spectral fittings with the continuum model {\em bbody+ power*highecut}, there were the absorption features in the energy range of $70 -90$ keV, which should be the cyclotron resonant scattering features. We have used the line model {\em cyclabs} to fit the cyclotron absorption line. In addition, around the energy of 6--7 keV, the iron line feature were also found in the residuals, thus we add the gaussian emission line to fit the Fe K$\alpha$ line.

In Figure 4, we showed the X-ray spectral examples of GRO J1008-57 in four outbursts. The all fitted parameters of these spectra are collected in Table 3. In the 2007 outburst, the significant absorption feature around 80 keV was confirmed, which was attributed to the cyclotron resonant scattering feature (also see Ge et al. 2020). This cyclotron resonant scattering feature was also detected during the peak epoch of the 2020 outburst. During the type I outbursts in 2018 and 2019, the cyclotron absorption lines around 70-90 keV cannot be detected with Insight-HXMT. The continuum spectral properties did not change with the different outbursts and luminosity levels.

\begin{figure*}
\includegraphics[width=15cm]{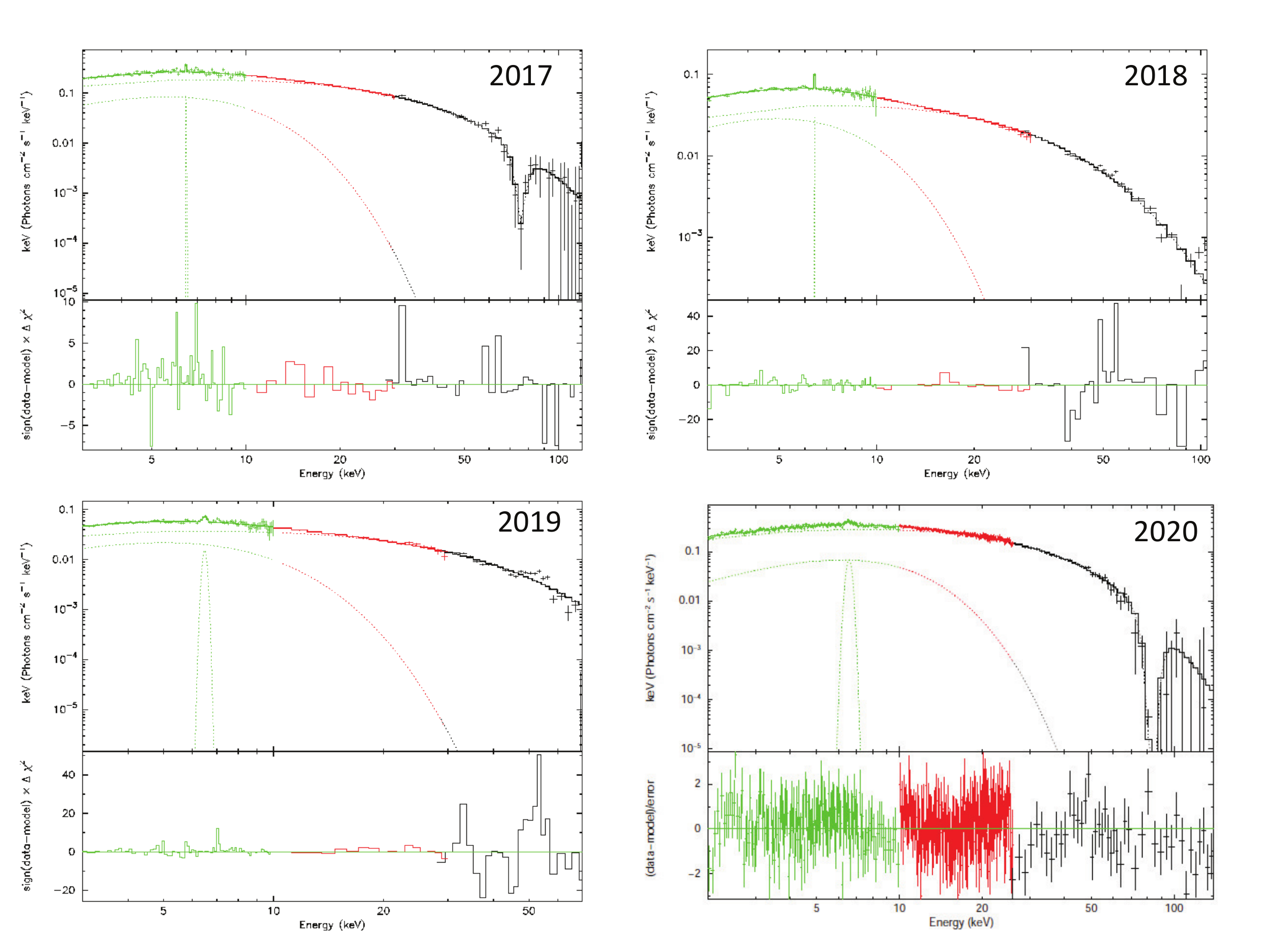}
\caption{The X-ray spectra of GRO J1008-57 obtained by Insight-HXMT in four outbursts: August 2017 (with the cyclotron absorption line), Feb 2018, June 2019, June 2020 (with the cyclotron absorption line). }
\end{figure*}

\begin{table}
	\caption{The X-ray spectral parameters of GRO J1008-57 in four outbursts. X-ray flux from 3--100 keV is given in units of erg cm$^{-2}$ s$^{-1}$. }
	\centering
\scriptsize
\begin{tabular}{@{}lllll@{}}
\hline
		
Parameter & 2017 & 2018 & 2019 & 2020 \\ \hline
kT (keV) & $1.99\pm 0.06$   & 1.74$\pm 0.02$   & 1.77$\pm 0.04$ & $2.11\pm 0.09$ \\
$\Gamma$ & $0.56\pm 0.06$ & $0.52\pm 0.02$ & $0.53\pm 0.07$ & 0.48$\pm 0.06$ \\
$E_{cut}$ (keV) & $5.58\pm 0.64$ & $6.01\pm 0.36$ & $4.41\pm 0.53$ & 4.21$\pm 0.63$ \\
$E_{f}$ (keV) & $16.56\pm 0.67$ & $15.33\pm 0.23$  & $14.01\pm 0.52$ & $14.58\pm 0.95$ \\
$E_{fe}$ (keV) & $6.46\pm 0.13$ & $6.46\pm 0.13$ & $6.48\pm 0.14$ & $6.55\pm 0.08$ \\
$\sigma$ (keV) & 0.08$\pm 0.06$ & 0.01 (fixed) & $0.09\pm 0.06$ & 0.17$\pm 0.06$ \\
$E_{cycl}$ (keV) & 76.6$\pm 1.9$ & - &  - & $82.4\pm 3.7$ \\
$W_{cycl}$ (keV) &4.3$\pm 2.1$ & - & - & $4.6\pm 2.8$\\
$D_{cycl}$ & $3.9\pm 1.1$ & - & - & $10.9\pm 4.3$ \\
Flux & 1.1$\times 10^{-8}$  & 2.3$\times 10^{-9}$ & 1.8$\times 10^{-9}$ & 1.0$\times 10^{-8}$\\
$\chi^2/d.o.f$ & 0.9896 & 1.1928 & 1.0351 & 0.9327 \\
\hline
\end{tabular}
\end{table}

We can calculated the value of the magnetic field of the neutron
star in GRO J1008-57 by using the formula \beq [B/10^{12}{\rm
G}]=[E_{\rm cycl}/11.6{\rm keV}](1+z), \enq where $E_{\rm cycl}$
is the energy of the fundamental line, here $E_{\rm cycl}=80$
keV, and $z$ is the gravitational redshift near the surface of the
neutron star. For a canonical neutron star of 1.4 \ms with a
radius of 10 km, we can find $z\sim 0.3$ (Kreykenbohm et al.
2004). Then we obtain a magnetic field of $\sim 9\times 10^{12}$ G for
the neutron star in GRO J1008-57.

\section{Discussion and conclusion}

With the Insight-HXMT data, we studied the temporal properties of the neutron star in GRO J1008-57 from August 2017 to June 2020, revealing the accretion torque reversals occurring around February 2018 and June 2020. As mentioned in the previous articles (e.g., Wang 2014), the neutron star in GRO J1008-57 showed a long time evolution of the spin period. GRO J1008-57 experienced a spin-down process from 1993 to 2009, and its mean spin-down rate was $(3.5\pm 0.3)\times 10^{-5}$ s/d, and it might change from spin-down to spin-up process after 2009 (Wang 2014). Combined with the work of the previous articles, the evolution of the spin period in GRO J1008-57 from 1993 to 2020 has be displayed in Fig. 5.

From the present observations, we confirmed the spin up trend of the neutron star after 2009, and inferred a spin-up rate of $ -(2.10\pm 0.05)\times 10^{-4}$ s/d from 2009 to 2018. Furthermore, after the type I outburst in Feb 2018, the accreting torque of the neutron star in GRO J1008-57 changed, the neutron star turned into the spin-down process, with a spin-down rate of $(6.7\pm 0.6)\times 10^{-5}$ s/d from 2018 to 2019. During the torque reversal, the pulse profiles of X-ray pulsar in GRO J1008-57 did not vary with different time. In addition, we also compared the X-ray spectra of GRO J1008-57, the continuum spectral properties of the outbursts from 2017 --2019 did not change significantly, while the cyclotron absorption line features can not be detected in two outbursts of 2018 and 2019. During the recent type II outburst in 2020, neutron star in GRO J1008-57 showed the fast spin-up process again with a rate of $\sim -9\times 10^{-4}$ s/d. The cyclotron line absorption feature around 80 keV was detected during the peak of the 2020 outburst.

\begin{figure}
	\includegraphics[width=9cm]{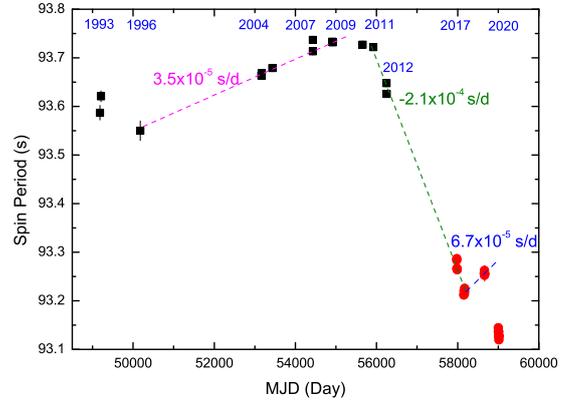}
	\caption{The spin period evolution of the neutron star in GRO J1008-57 from 1993 to 2020 collected from different observations including the new Insight-HXMT observations. }
	
\end{figure}

{\em Spin up process from 2009 -- 2018} \\
Interactions between the strongly magnetized neutron stars and surrounding accretion disk are thought to be the dominant mechanism to drive the spin evolutions of the neutron stars. The classical torque model proposed by Ghosh \& Lamb (1979) suggested that the magnetic field lines were threaded in the Keplerian accretion disk in a broad transition zone, which will derive the torque acting on the neutron star. The torque from the inner accretion matter flow into the neutron star will drive the spin up process. The magnetic torque due to the magnetosphere interacting with the matter will result in the spin down of the neutron star. The positive or negative torque would depend on a so-called {\em fastness parameter} (Ghosh \& Lamb 1979, Wang 1995): $\omega=\Omega_{NS}/\Omega_0 = (R_0/R_c)^{3/2}$, where $\Omega_{NS}$ is the angular velocity of the star and $\Omega_0$ is the the Keplerian angular velocity at the inner radius $R_0$ of the disk. The inner radius $R_0$ is generally thought to be in the same order of the Alfv\'{e}n radius $R_A$, here they can be expressed in the form of $R_0 =\xi R_A$, where $R_A=\mu^{4/7} {\dot M}^{-2/7}(2GM)^{-1/7}$ in the case of spherical
accretion, and $\mu=BR^3$, $B$ is the surface dipole magnetic field of neutron star, $M$ is the mass of neutron star, $\dot M$ is the accretion rate.

The neutron star of GRO J1008-57 undergone the long-term spin-up process from 2009 to 2018. Based on the accreting torque model and the observed spin-up rate, we could estimate the surface magnetic field of GRO J1008-57. The spin-up rate during the accreting state can be expressed as (Joss \& Rappaport 1984; Wang 1996):
\beq  -{\dot P \over P}= \dot M (GM R_0)^{1/2} {n(\omega) \over I\Omega_{NS}}, \enq
where $I\sim 2MR_{NS}^2/5$ is the neutron star's moment of inertia, then we find
\beq
-{\dot P \over P}\simeq 1.35\times 10^{-12}{\rm s^{-1}} \omega^{1/3} n(\omega) L_{37} P^{4/3} ({M \over 1.4M_\odot})^{-4/3}R_6^{-1}.
\enq
$L_{37}$ is the X-ray luminosity of the X-ray pulsar in units of $10^{37}$ erg cm$^{-2}$ s$^{-1}$, $R_6$ is the radius of the neutron star in units of $10^6$ cm. $n_{\omega}$ is a dimensionless torque function given by (Wang 1995):
\beq n(\omega)= {(7/6)-(4/3)\omega+(1/9)\omega^2 \over 1-\omega} .
\enq
Thus we can deduce the dipole magnetic field moment of the star:
\beq
\mu_{30} \sim 0.25\xi^{-7/4}\omega^{7/6}L_{37}^{1/2}P^{7/6}({M\over 1.4M_\odot})^{1/3}R_6^{1/2},
\enq
where $\mu_{30}$ is the dipole moment of the star in units of $10^{30}$ G cm$^3$.

In the case of GRO J1008-57, the spin period of the neutron star is 93.3 s. The mean accretion rate is uncertain. The X-ray luminosities in several type II outbursts (e.g. March, 2009, November 2012, November 2014, January 2015, August 2017) are in the range of $\sim 10^{37} - 2\times 10^{38}$ erg s$^{-1}$ (Wang 2014; Yamamoto et al. 2014; Bellm et al. 2014; Ge et al. 2020). More frequent Type I bursts have the mean luminosity around $10^{37}$ erg/s (K\"uhnel et al. 2017). For the quiescent state, the system has a mean luminosity of $\sim 10^{35}$ erg/s or higher (Tsygankov et al. 2017). During the spin-up stage from 2009 - 2017, we take the mean accretion luminosity of $10^{37}$ erg/s.  Then we derived the "fastness parameter" $\omega \simeq 0.95$, furthermore determined the surface magnetic field of the neutron star in GRO J1008-57 as $B\simeq 4\times 10^{13} \xi^{-7/4}$ G. If the inner radius of the accretion disk is near the Alfv\'{e}n radius, i.e., $\xi\sim 1$, one will derive the strong magnetic field in GRO J1008-57, which is higher than that obtained from the cyclotron line measurement. The similar results were obtained by Shi et al. (2015) that the surface magnetic fields of most neutron stars in known Be/X-ray pulsars are higher than 10$^{14}$ G estimated by spin equilibrium using the accretion torque and magnetosphere models (Dai \& Li 2006).

There still exist some uncertainties in estimating the magnetic field, like the inner radius of the accretion disk. If we assume that the magnetic field determined by the cyclotron line centroid energy is the real value of the surface magnetic field in GRO J1008-57, then we find $\xi \sim 2.2$. Thus the inner radius of the accretion disk is larger than the Alfv\'{e}n radius in the case of GRO J1008-57. It could be possible for X-ray pulsars in Be/X-ray binaries. In some Be/X-ray pulsars, the quasi-periodic oscillations were observed (Roy et al. 2019, Takeshima et al. 1994, and references therein). Using the beat frequency model, Takeshima et al. (1994) directly determined the inner radius of these X-ray pulsar which was about 1 --3.5 times of the Alfv\'{e}n radius.

Here it should be pointed out that we have used the torque function based on the work by Wang (1995). This model predicts the critical fastness parameter in the range of 0.88 - 0.95 near the spin equilibrium. Some other models give the lower values of the critical fastness parameter, e.g., 0.7 - 0.9 suggested by Li \& Wang (1996, 1999), 0.4 - 0.6 by Li \& Wickramasinghe (1997), and $\sim 0.35$ based on Ghosh \& Lamb (1979). If we consider the smaller values of the critical fastness parameter (i.e.,$\omega\sim 0.35$), then $B\sim 1.1\times 10^{13} \xi^{-7/4}$ G. Thus by comparing the magnetic field value from the CRSF measurement, we can find $\xi\sim 1$, the inner disk radius would be close to the Alfv\'{e}n radius in GRO J1008-57, which is also consistent with the conclusion by Li \& Wang (1999).  

{\em Spin down process from 2018 -- 2019} \\
After February 2018, the X-ray pulsar in GRO J1008-57 transferred to the spin-down process. Similar to the long-term spin-down trend observed from 1993-- 2009, these spin down processes might be the evidence for the propeller phase in which the angular momentum of the neutron star is removed by the interaction between matter and the magnetosphere (Illarionov \& Sunyaev 1975). The other possibility is that the neutron star has not gone into the propeller phase yet, but the spin-up torque from the accretion flows is smaller than the magnetic torque, then the neutron star in GRO J1008-57 showed the spin-down behavior. In Fig. 3, the pulse X-ray emissions in GRO J1008-57 during 2018 -- 2019 can be clearly detected, suggesting that the accretion still occurred on the surface of neutron star. Thus we think that the latter possibility may better describe the spin-down process in GRO J1008-57.

If we neglected the spin-up torque from accretion flow during the spin-down process, we can estimate the low limit of the magnetic field strength of the neutron star. The magnetic torque should be a little larger than the observed spin-down torque of the system, i.e., \beq \mu^2/R_c^3 \geq I|\dot \omega|, \enq where $R_c=(GMP^2/4\pi^2)^{1/3}$ is the corotation radius of the neutron star. Given the observed mean spin-down rate from 2018 -- 2019, $\dot P=6.7\times 10^{-5}$ s day$^{-1}$, the derived surface magnetic field is $B\geq 6\times 10^{12}$ G, which is consistent with the strength measured by the cyclotron absorption lines (this work, also see Ge et al. 2020).

{\bf Acknowledgements} \\
We are grateful to the referee for the suggestions to improve the manuscript. This work made use of the data from the Insight-HXMT mission, a project funded by China National Space Administration
(CNSA) and the Chinese Academy of Sciences (CAS). The authors thank the support by the National Program on Key Research
and Development Project (Grants No. 2016YFA0400803) and the NSFC (U1838103, 11622326, U1838201 and U1838202).

\end{document}